\documentclass[twocolumn,amsmath,amssymb,10pt,superscriptaddress,a4paper,letterpaper,fleqn]{revtex4-1}
\usepackage{amssymb}
\usepackage{epsfig}
\usepackage{graphicx}
\usepackage{dcolumn}
\usepackage{array}
\usepackage{bm}
\usepackage{fancyheadings}
\usepackage{longtable}
\usepackage{multirow}
\usepackage{float}
\usepackage{afterpage}
\usepackage{color}

\newcommand{\be}{\begin{equation}}
\newcommand{\ee}{\end{equation}}
\newcommand{\ba}{\begin{eqnarray}}
\newcommand{\ea}{\end{eqnarray}}

\newcommand{\Tr}{{\rm Tr}}

\setlongtables
\usepackage[breaklinks=true,linkbordercolor={1 1 1}]{hyperref}

\begin{document}
\setcounter{page}{1}


\title{
\qquad \\ \qquad \\ \qquad \\  \qquad \\  \qquad \\ \qquad \\
Level Densities of Heavy Nuclei by the Shell Model Monte Carlo Method}

\author{Y. Alhassid}
\email[Corresponding author:]{yoram.alhassid@yale.edu}
\affiliation{Center for Theoretical Physics, Sloane Physics Laboratory,Yale University, New Haven, CT 06520, USA}

\author{C. \"Ozen}
\affiliation{Faculty of Engineering and Natural Sciences, Kadir Has University, Istanbul 34083, Turkey}

\author{H. Nakada}
\affiliation{Department of Physics, Graduate School of Science,Chiba University, Inage, Chiba 263-8522, Japan}

\date{\today}

\begin{abstract}
{The microscopic calculation of nuclear level densities in the presence of correlations is a difficult many-body problem. The shell model Monte Carlo method provides a powerful technique to carry out such calculations using the framework of the configuration-interaction shell model in spaces that are many orders of magnitude larger than spaces that can be treated by conventional methods. We present recent applications of the method to the calculation of level densities and their collective enhancement factors in heavy nuclei. The calculated level densities are in close agreement with experimental data.
}
\end{abstract}
\maketitle

\lhead{ND 2013 Article $\dots$}
\chead{NUCLEAR DATA SHEETS}
\rhead{Y. Alhassid \textit{et al.}}
\lfoot{}
\rfoot{}
\renewcommand{\footrulewidth}{0.4pt}

\section{ INTRODUCTION}

Level densities are an integral part of the Hauser-Feshbach theory~\cite{Hauser1952} of statistical nuclear reactions and appear in numerous nuclear physics applications. However, in the presence of correlations, their microscopic calculation is a challenging many-body problem. The configuration-interaction (CI) shell model method is a suitable framework to calculate level densities in that it accounts for shell effects and correlations, but the dimension of the many-particle Hilbert space increases combinatorially with the number of nucleons and/or the number of single-particle orbitals, prohibiting such calculations in mid-mass and heavy nuclei. This difficulty can be overcome by using the shell model Monte Carlo (SMMC) method~\cite{Lang1993,Alhassid1994,Koonin1997,Alhassid2001}. This method enables calculations of statistical and collective properties of nuclei in very large model spaces and proved to be a powerful technique in the microscopic calculation of level densities~\cite{SMMC-densities}.

Here we discuss recent applications of SMMC to heavy rare-earth nuclei and, in particular, to isotopic families of samarium and neodymium nuclei~\cite{Ozen2013a,Ozen2013b}. We demonstrate that with a proper choice of the model space, the CI shell model approach is capable of describing microscopically the crossover from vibrational-to-rotational collectivity in these nuclei as the number of neutrons increases from shell closure towards mid-shell. We calculate the SMMC state densities for both the even- and odd-mass samarium and neodymium isotopes, and find them to be in very good agreement with experimental data. Defining the collective enhancement factor as the ratio of the SMMC state density to the Hartree-Fock-Bogoliubov (HFB) density, we calculate microscopically these collective enhancement factors in the even samarium and neodymium isotopes and study their decay with excitation energy.

\section{THE SHELL MODEL MONTE CARLO METHOD}

The SMMC method is based on a representation of the Gibbs operator $e^{-\beta H}$ ($\beta=1/T$ is the inverse temperature and $H$ is the CI shell model Hamiltonian) as a superposition of one-body propagators of non-interacting nucleons moving in external time-dependent auxiliary fields $\sigma = \sigma(\tau)$. This representation, known as the Hubbard-Stratonovich transformation, is given by
\be\label{HS}
e^{-\beta H} = \int D[\sigma] G_\sigma U_\sigma \;,
\ee
where $G_\sigma$ is a Gaussian weight and $U_\sigma$ is a one-body propagator for a configuration $\sigma$ of the auxiliary fields. The thermal expectation value of an observable $O$ is given by
\be \label{observable}
\langle O\rangle = {\Tr \,( O e^{-\beta H})\over  \Tr\, (e^{-\beta H})} = {\int D[\sigma] W_\sigma \Phi_\sigma \langle O \rangle_\sigma
\over \int D[\sigma] W_\sigma \Phi_\sigma} \;,
\ee
where $W_\sigma = G_\sigma |\Tr\, U_\sigma|$ is a positive-definite function, $\Phi_\sigma = \Tr\, U_\sigma/|\Tr\, U_\sigma|$ is known as the Monte Carlo sign function, and $\langle O \rangle_\sigma = \Tr \,(O U_\sigma)/ \Tr\,U_\sigma$.  Here the traces are evaluated in the canonical ensemble, i.e., at fixed numbers of protons and neutrons.  In SMMC, auxiliary-field configurations $\sigma_k$ are chosen according to $W_\sigma$, and the expectation value in (\ref{observable}) is then estimated from
 $\langle  O\rangle \approx  {\sum_k
  \langle O \rangle_{\sigma_k} \Phi_{\sigma_k} / \sum_k \Phi_{\sigma_k}}$.

\section{HEAVY NUCLEI}

The SMMC method was extended to heavy nuclei, overcoming a number of technical challenges~\cite{Alhassid2008}. While even-even mid-mass nuclei are characterize by small deformation and a first excitation energy of $\sim 1-2$ MeV, the situation is very different in even-even heavy nuclei, which can have large deformation and a much smaller first excitation energy of $\sim 100$ keV. As a result it necessary to propagate to much longer imaginary time $\beta$ to reach the ground state. The matrix representing $U_\sigma$ in the single-particle space becomes ill-condition at large $\beta$ and matrix multiplication has to be stabilized.

\subsection{Model Space and Hamiltonian}

The required model space in heavy nuclei is significantly larger in comparison with the model space used in mid-mass nuclei. For rare-earth nuclei we used the many-particle model space spanned by the single-particle orbitals $0g_{7/2}$, $1d_{5/2}$, $1d_{3/2}$, $2s_{1/2}$, $0h_{11/2}$ and $1f_{7/2}$ for protons, and the orbitals $0h_{11/2}$, $0h_{9/2}$, $1f_{7/2}$, $1f_{5/2}$,
$2p_{3/2}$, $2p_{1/2}$, $0i_{13/2}$, and $1g_{9/2}$ for neutrons. This corresponds to the $50-82$ major shell plus the $1f_{7/2}$ orbital for protons, and the $80-126$ major shell plus the $0h_{11/2}$ and $1g_{9/2}$ orbitals for neutrons. The calculations are done using an SMMC code in the proton-neutron formalism~\cite{Alhassid2008}.

We used the Hamiltonian of Ref.~\onlinecite{Ozen2013a}. The bare single-particle energies are taken so as to reproduce the Woods-Saxon energies in the spherical Hartree-Fock approximation. The interaction includes monopole proton and neutron pairing terms and several multipole-multipole interaction terms (quadrupole, octupole and hexadecupole) with coupling constants given in Ref.~\onlinecite{Ozen2013a}.

\subsection{Emergence of Collectivity}

Heavy nuclei are known to exhibit various types of collectivity, such as vibrational and rotational collectivity, that are well described by phenomenological models. However, a microscopic description in the framework of the CI shell model has been mostly lacking. Heavy nuclei require the use of a large model space that is prohibitively large for conventional diagonalization methods. The SMMC approach, while capable of treating much larger model spaces, is suitable for calculating thermal and ground-state observables but not for calculating detailed level schemes, often used to identify the specific type of collectivity.

\begin{figure}[t!]
\includegraphics[clip,width= 0.85\columnwidth]{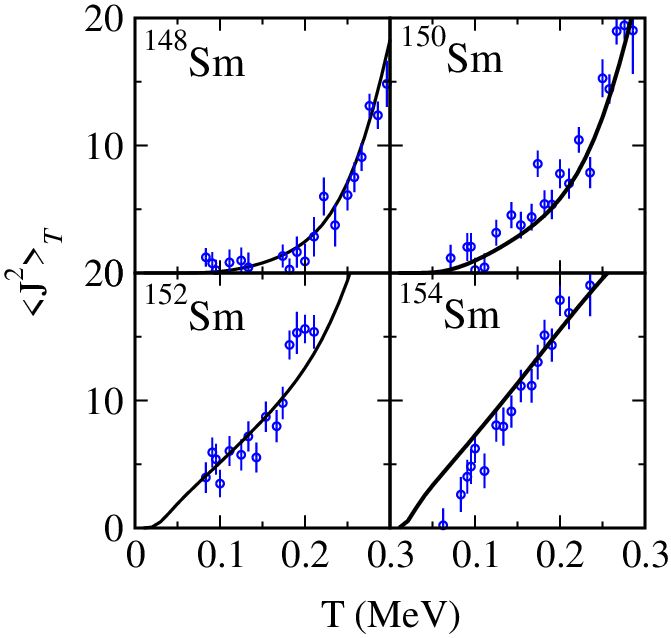}
\caption{$\langle \mathbf{J}^2 \rangle_T$ as a function of temperature in the even-even $^{148-154}$Sm isotopes. We compare the SMMC results (open circles with error bars) with the results extracted from the experiment (solid lines, see text). Adapted from Ref.~\onlinecite{Ozen2013a}. }
\label{J2}
\end{figure}

We were able to distinguish between the different types of collectivity by using the thermal observable $\langle \mathbf{J}^2 \rangle_T$ (where $\bf J$ is the total nuclear spin), whose low-temperature behavior is sensitive to the type of collectivity.
In Fig.~\ref{J2} we show  $\langle \mathbf{J}^2 \rangle_T$ as a function of temperature $T$ for a family of even-mass samarium isotopes. The SMMC results (open circles) display a crossover from a ``soft'' response to temperature in $^{148}$Sm, characteristic of a vibrational nucleus, to a ``rigid'' response in $^{154}$Sm, typical of a rotational nucleus. The solid lines in Fig.~\ref{J2} are calculated from experimental data using a complete level scheme below a certain threshold energy and an empirical backshifted Bethe formula (BBF) for the level density above this threshold. The parameters of the BBF are determined by a fit to the cumulative counting data at low excitation energies and the neutron resonance data at the neutron separation energy. Neutron resonance data are unavailable for $^{154}$Sm, and the corresponding solid line in Fig.~\ref{J2} is obtained using discrete experimental levels only.

\subsection{State Densities}

\begin{figure}[t!]
\includegraphics[clip,width= 0.9\columnwidth]{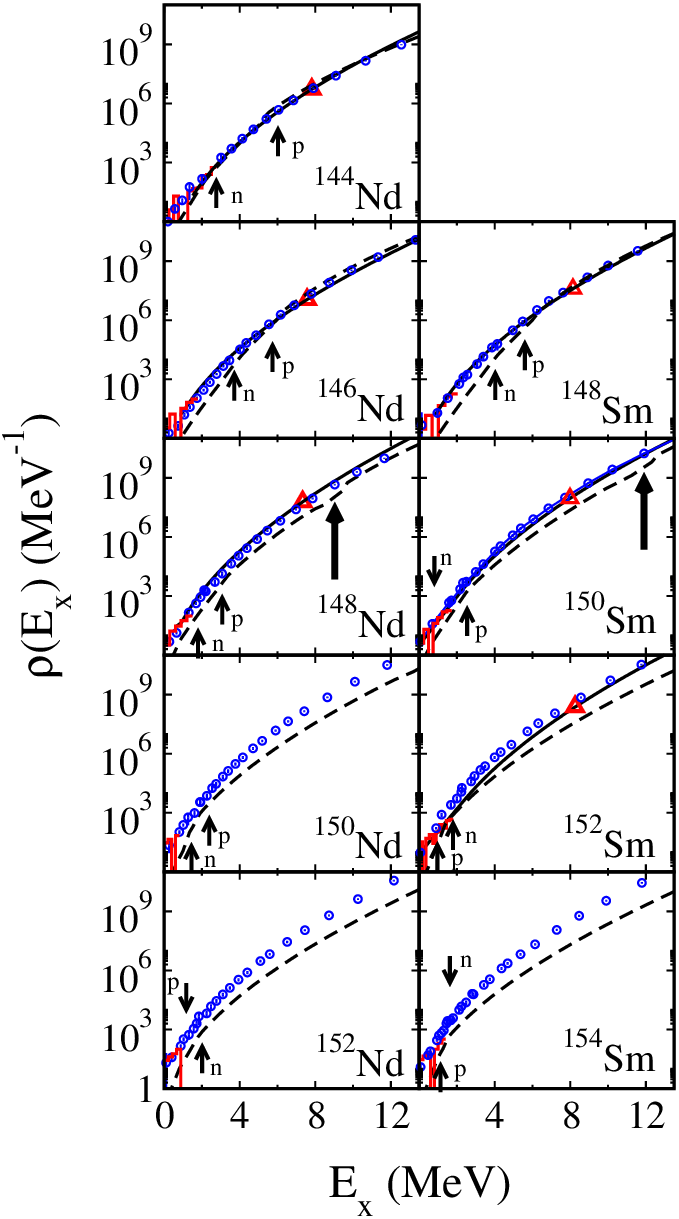}
\caption{State densities of even-mass $^{144-152}$Nd (left column) and  $^{148-154}$Sm  isotopes (right column). The SMMC densities (open circles) are compared with level counting data (histograms) and neutron resonance data at the neutron separation energy (open triangles). The solid lines are BBF densities fitted to the cumulative level counting data and neutron resonance data. The samarium results are adapted from Ref.~\onlinecite{Ozen2013a}.}
\label{rho_even}
\end{figure}

\begin{figure}[h!]
\includegraphics[clip,width=0.85\columnwidth]{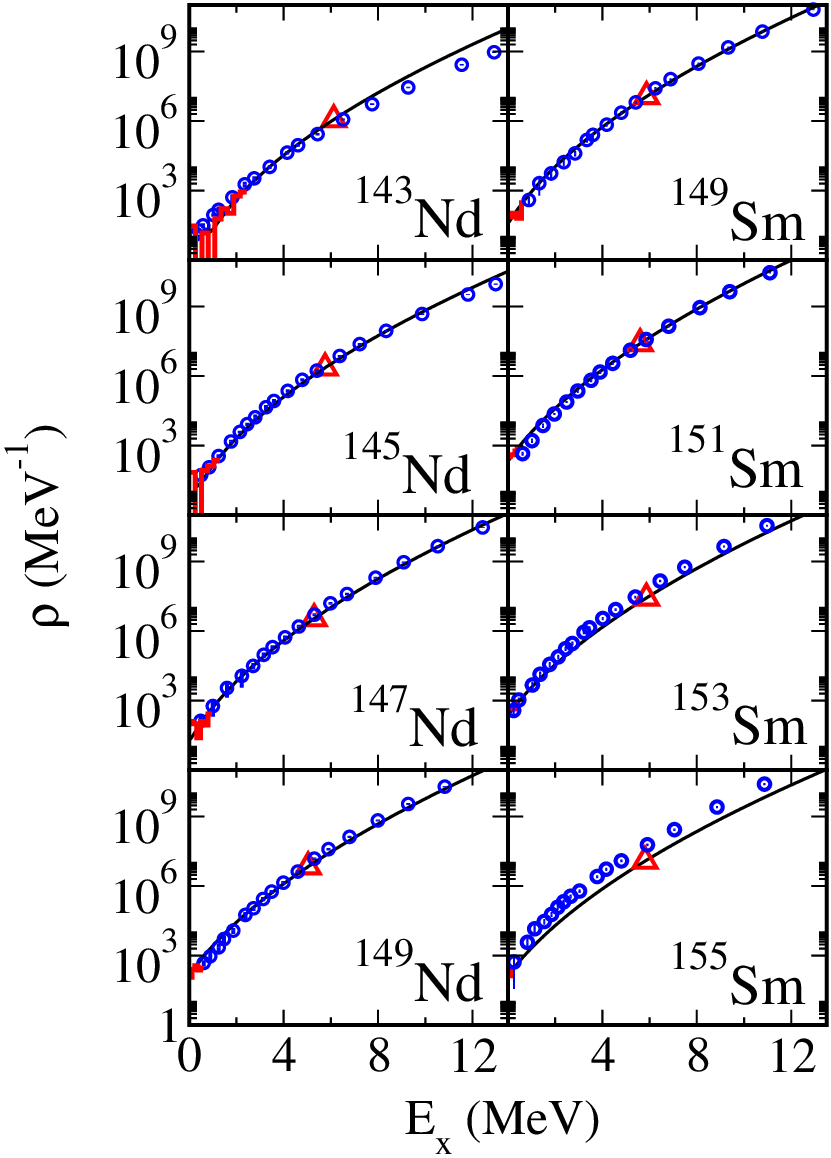}
\caption{State densities of odd-mass $^{143-149}$Nd (left column) and $^{149-155}$Sm (right column) isotopes. Symbols and lines are as in Fig.~\ref{rho_even}. Adapted from Ref.~\onlinecite{Ozen2013b}.}
\label{rho_odd}
\end{figure}

The state density is the inverse Laplace transform of the partition function $Z(\beta) = \Tr  e^{-\beta H}$. The average state density is found by evaluating this integral transform in the saddle-point approximation. We have~\cite{Ericson1960}
\begin{equation}
\label{eq:levdens}
\rho(E) \approx \frac{1}{\sqrt{2 \pi T^2 C}} e^{S(E)} \;,
\end{equation}
where  $S(E)$ and $C$ are, respectively, the entropy and heat capacity in the canonical ensemble. In SMMC, we calculate the thermal energy $E(\beta)= \langle H \rangle$ as a thermal observable and then integrate the thermodynamic relation $-d\mathrm{ln} Z/d\beta = E(\beta)$ to find $Z(\beta)$. The entropy and heat capacity are then calculated from
\begin{equation}
S(E)=\ln Z+\beta E\;;\;\;\;\;\; C=-\beta^2 \partial E/\partial\beta \;.
\end{equation}

To compare with experiment, it is necessary to determine the state density as a function of excitation energy $E_x=E-E_0$, where $E_0$ is the ground-state energy. This requires an accurate estimate of $E_0$.

\subsubsection{Even-Even Nuclei}

The projection on an even number of particles for a good-sign interaction keeps the Monte Carlo sign good for almost all samples. Thus in even-even nuclei, we can carry out accurate SMMC calculations of the thermal energy $E(\beta)$ up to large $\beta$ values and use them to obtain a reliable estimate of the ground-state energy $E_0$.

In Fig.~\ref{rho_even} we show the calculated SMMC state densities (open circles) for the even-even $^{144-152}$Nd and $^{148-154}$Sm isotopes. The results are in good agreement with level counting data (histograms) and, when available, neutron resonance data at the neutron separation energy (open triangles). The dashed lines are the densities calculated in the HFB approximation. The kinks in the HFB density correspond to the neutron and proton pairing phase transitions (arrows) and to a shape phase transition in nuclei that are deformed in their ground state (thick arrows).

\subsubsection{Odd-Even Nuclei}

The projection on an odd number of particles leads to a new sign problem (even for good-sign interactions) at low temperatures. As a result, the thermal energy for the odd-even rare-earth nuclei can be calculated  in practice only up to $\beta \sim 4 - 5$ MeV$^{-1}$. It is therefore not possible to obtain a reliable estimates for the ground-state energies of odd-even nuclei in direct SMMC calculations.

To extract a ground-state energy $E_0$ for an odd-even nucleus, we carried out a one-parameter fit of the SMMC thermal excitation energy $E_x(T)=E(T)-E_0$ to the experimental thermal energy. The latter is obtained from $E_x(\beta)=-d\ln Z(\beta)/d\beta$, where $Z$ is the partition function extracted from experimental data using a complete level scheme at low excitations below a certain threshold energy and an experimentally determined BBF level density above it. In Fig.~\ref{rho_odd} we show the state densities of odd-mass neodymium and samarium isotopes. The SMMC state densities are in close agreement with experimental data, except for some discrepancies with the neutron resonance data in $^{153}$Sm and $^{155}$Sm.

\begin{figure}[t!]
\includegraphics[clip,width=0.9\columnwidth]{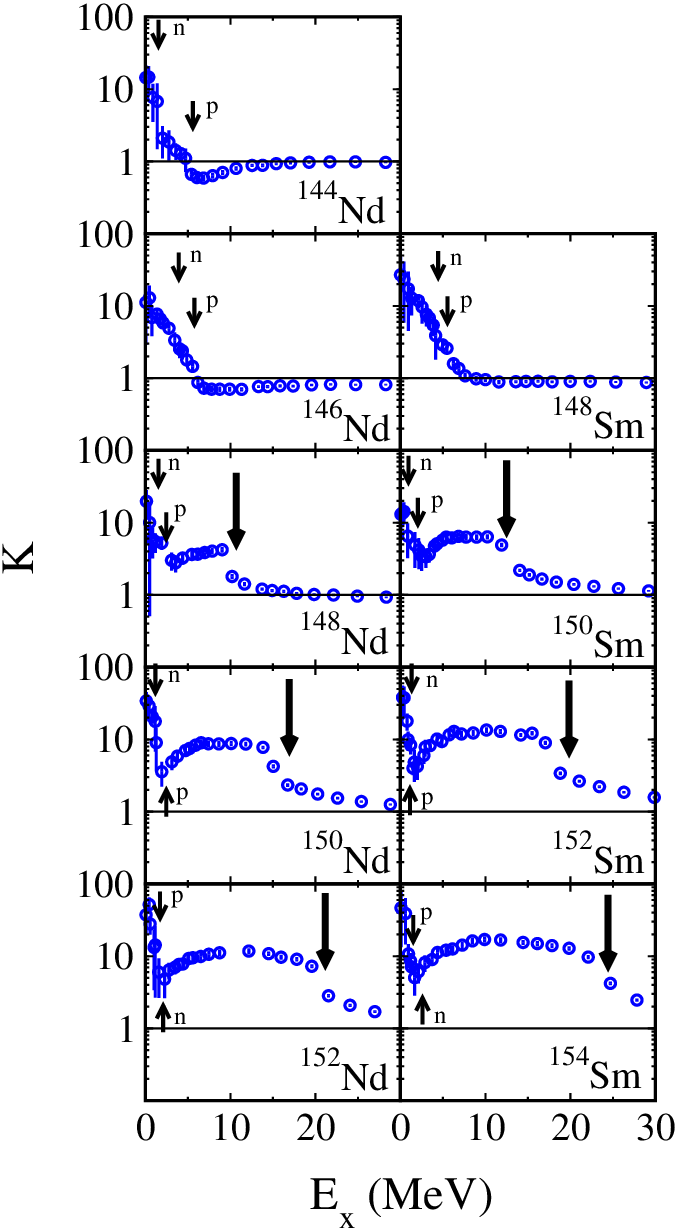}
\caption{Collective enhancement factor $K$ (open circles) of even-mass neodymium (left column) and samarium (right column) isotopes versus excitation energy $E_x$. Arrows are  as in Fig.~\ref{rho_even}. The samarium results are adapted from Ref.~\onlinecite{Ozen2013a}.}
\label{enh}
\end{figure}

\subsection{Collective Enhancement Factors}

The enhancement of level densities by collective states is described by the collective enhancement factors. Their decay with excitation energy is one of the least understood topics in the modeling of level densities~\cite{RIPL}. We define a collective enhancement factor $K$ to be the ratio between the SMMC state density and the HFB density, $K = \rho_{\rm SMMC}/\rho_{\rm HFB}$. The HFB density describes the density of intrinsic states, and the enhancement observed in the SMMC  density originates in rotational bands that are built on top of the intrinsic states and in vibrational collectivity that is missed in the HFB approximation.

In Fig.~\ref{enh} we show $K$ for isotopic families of neodymium and samarium nuclei. In the spherical nuclei $^{144}$Nd, $^{146}$Nd and $^{148}$Sm the enhancement $K$ is due to vibrational collectivity, and we observe the decay of $K$ to $\sim 1$ above the neutron and proton pairing phase transition energies. However, in deformed nuclei $K$ has a minimum above the pairing transition energies, and its rise at higher excitations is due to rotational collectivity. In these nuclei, $K$ decays to $\sim 1$ above the shape phase transition energy.

\section{ CONCLUSION}

We used the SMMC method to calculate state densities of heavy rare-earth nuclei in the CI shell model framework and found close agreement with experimental data. We also calculated the collective enhancement factors of state densities and observed that the decays of vibrational and rotational collectivity correlate with the pairing and shape phase transitions, respectively.

This work is supported in part by the U.S. DOE grant
No. DE-FG-0291-ER-40608 and as Grant-in-Aid for Scientific Research (C) No. 25400245 by the JSPS, Japan.  Computational cycles were provided by the NERSC high performance computing facility at LBL and by the facilities of the Yale University Faculty of Arts and Sciences High Performance Computing Center.


\begin{thebibliography}{9}

\bibitem{Hauser1952} W.~Hauser and H.~Feshbach, {\sc Phys. Rev.} {\bf 87}, 366 (1952).
\bibitem{Lang1993} G.H.~Lang,  C.W.~Johnson, S.E.~Koonin, and W.E.~Ormand, {\sc Phys. Rev. C} {\bf 48}, 1518 (1993).
\bibitem{Alhassid1994} Y.~Alhassid {\it et al.} {\sc Phys. Rev. Lett.}, {\bf 72}, 613 (1994).
\bibitem{Koonin1997} S.E. Koonin, D.J. Dean, and K. Langanke, {\sc Phys. Rep.} {{\bf 278}}, 2 (1997).
\bibitem{Alhassid2001} Y.~Alhassid, {\sc Int. J. Mod. Phys.} B {\bf 15},  1447 (2001).
\bibitem{SMMC-densities} H. Nakada and Y. Alhassid, {\sc Phys. Rev. Lett.} {\bf 79}, 2939 (1997);
 K. Langanke, {\sc Phys. Lett.} B {\bf 438}, 235 (1998);
 Y. Alhassid, S. Liu and H. Nakada, {\sc Phys. Rev. Lett.} {\bf 83}, 4265 (1999); {\it ibid.} {\bf 99}, 162504 (2007);  C. \"{O}zen, K.~Langanke, G.~Martinez-Pinedo, and D.~J.~Dean, Phys. Rev. C {\bf 75} 064307 (2007).
\bibitem{Ozen2013a} C.~\"{O}zen, Y.~Alhassid, and H. Nakada, {\sc Phys. Rev. Lett.} {\bf 110}, 042502 (2013).
\bibitem{Ozen2013b} C.~\"{O}zen, Y.~Alhassid, and H. Nakada, arXiv:1304.7405.
\bibitem{Alhassid2008} Y. Alhassid, L. Fang and H. Nakada, {\sc Phys. Rev. Lett.} {\bf 101}, 082501 (2008).
\bibitem{Ericson1960} T. Ericson, {\sc Adv. Phys.} {\bf 9}, 425 (1960).
\bibitem{RIPL} R. Capote {\it et al.}, {\sc Nuclear Data Sheets} {\bf 110}, 3107 (2009).
\end{thebibliography}
\end{document}